# *surfQuake: A new Python toolbox for the workflow process of seismic sources*

Roberto Cabieces[1], Thiago C. Junqueira[2], Katrina Harris[3], Jesús Relinque[1], Claudio Satriano[4].

1. Department of Geophysics, Spanish Navy Observatory, San Fernando, Spain.

2. Institute of Geoscience, University of Potsdam, Karl-Liebknecht-Str., 24–2514476 Potsdam-Golm, Germany.

3. Department of Earth Sciences, University College London, Gower Place, WC1H6BT London, UK.

4. Institut de Physique du Globe de Paris (IPGP), 1 Rue Jussieu 75238 Paris cedex 05, France

**Abstract**

surfQuake is a new software designed to streamline the estimation of seismic source parameters. Its comprehensive set of toolboxes automate the determination of seismic arrival times, event association and locations, moment magnitude from P- or S- wave displacement spectra and moment tensor inversions within a Bayesian framework.

surfQuake is programmed in Python 3 and offers the users the possibility of three programming levels for flexibility and customization. The core library allows users to integrate the core of surfQuake into their preexisting scripts, giving advanced users full control, while the Command Line Interface gives users access to an upper layer that simplifies the use of the core. Alternatively, surfQuake core is wrapped by a Graphical User Interface (GUI) and connected to a SQLite database making it accessible to users with little coding experience.

The software has been fully tested with an earthquake cluster of more than 2000 events, that occurred in central Pyrenees in 2021-22. The source parameters retrieved from the cluster and the basic statistics associated with them are displayed using the surfQuake database toolbox. Additionally, we offer a web tutorial with the documentation of surfQuake and a set of usage examples for the three programming levels.

1. **Introduction**

In the last few years there have been multiple efforts to develop open-source software tools to process large seismic data sets, such as *Integrated Seismic Program*, (Cabieces et al., 2022) and *SEISAN* (Havskov et al., 2020). Specific seismic libraries including *ObsPy* (Krischer et al. 2015), *Pyrocko* (Heimann et al. 2019), *Computer Programs in Seismology* (Herrmann 2013) or *Seismic Analysis Code* (Goldstein et al., 2003) have already helped researchers to develop new tools to manage seismic data, making it easier to face new challenges in seismology.

Recent advancements in deep machine learning have automated routine seismic tasks, such as accurately determining the arrival time of seismic phases (Ross et al. 2018; Zhu y Beroza 2019; Mousavi et al., 2020) and associating them to corresponding events (M. Zhang et al., 2019; X. Zhang et al., 2021; Zhu et al., 2022; Münchmeyer, 2024). These new advancements have opened up the possibility of establishing a procedural workflow for automatically determining event source parameters from large datasets.

Several recent projects have developed automated workflows for monitoring earthquakes (Si et al., 2024; Wu et al., 2022; Zhang et al., 2022). However, there remained an opportunity to combine an event location workflow and a reliable estimation of source parameters. In addition, previous software has been written assuming previous knowledge of programming, preventing wider usage.

In this paper, we present a new Python 3 open-source software. surfQuake comprises a suite of toolboxes designed to automate the process of determining event source parameters. It encompasses the various stages of this procedure, including seismic arrival time determination facilitated by deep neural network techniques (Zhu & Beroza 2019), event association (Zhang et al., 2021), estimation of source parameters through the inversion of body-wave displacement spectra (Satriano, 2023), and determination of the focal mechanism using a moment tensor inversion approach within a Bayesian framework (Vackář et al. 2017). The automation of monitoring and determination of seismic source parameters proves especially valuable in scenarios where real-time data acquisition is unfeasible, as observed in Ocean Bottom Deployments (Cabieces et al., 2020a, 2020b). In such instances, large datasets are stored for subsequent processing, presenting a laborious challenge for technicians and researchers.

The design of surfQuake offers users the possibility of working at three levels. The first level is a dedicated Graphical User Interface (GUI) intended for users with little programming experience who prioritize interactivity rather than automatization or high degree of flexibility. The second level is a Command Line Interface (CLI) which offers a mid-point between the interactivity of the GUI and full flexibility. The last level is the surfQuake Core Library (CL) which allows full flexibility and integration to the user's scripts or applications. The design permits an easy scalability of the Core Library, enabling the implementation of new functionalities and their integration to the CLI and GUI. Following this approach ensures that surfQuake has the flexibility to incorporate new algorithms and neural networks.

Another important feature of surfQuake is its compatibility with other seismic software that offer users the ability to interactively explore seismic data (*Integrated Seismic Program*, ([Cabieces et al., 2022](#)) or *SEISAN* ([Havskov et al., 2020](#))). The results obtained from surfQuake, such as seismic catalogs or databases, can be conveniently accessed through ISP and *SEISAN*, respectively. Additionally, users can review seismic waveforms associated with a specific project across the programs.

surfQuake has been tested with one year of seismic data from ten stations in the Pyrenees region. The earthquake cluster located in the study region is displayed in Figure 1 from the Institut Cartogràfic i Geològic de Catalunya (ICGC) seismic catalog. The seismic activity within this area is extensively monitored by a dense coverage of permanent land stations, enabling a robust evaluation of both detectability capabilities and the accuracy of estimated source parameters.

surfQuake is maintained by the Department of Geophysics at the Spanish Navy Observatory and is hosted in a GitHub repository (see Data and Resources) to allow users access to version upgrades. Users can directly contact the developers through this GitHub repository. surfQuake documentation is available online and is maintained by the developer team (see Data and Resources). The documentation describes the core library and provides detailed instructions on the usage of the CL, CLI and GUI.

surfQuake is a free open-software licensed under the GNU Lesser General Public License version 3.0 (v3.0).

## 2. Data

We have tested surfQuake with one year of seismic data (01-10-2021 to 01-10-2022) from the combined the Institut Cartogràfic i Geològic de Catalunya (CA/ICGC: https://doi.org/10.7914/SN/CA) and the French Seismological and Geodetic (FR/RESIF: https://doi.org/10.17616/R37Q06) network. Detailed information about the seismic sensors is shown in Table 1 of the Supplementary Material.

The data was acquired from 10 broad-band seismometers located in the Central Pyrenees. This dataset provided an excellent scenario to test the software. The high seismicity of 2000 events of moderate magnitude (M < 3.8) facilitates a robust analysis of the detection and association capabilities, and the excellent seismic station coverage around the epicenters allow the evaluation of the event location and seismic parameters estimation.

Figure 1 displays the stations selected from the CA and FR seismic networks and the seismicity detected in the region of study (ICGC seismic catalog. A data subset from the highest seismicity period in the year has been included in a dedicated repository (see Supplementary Material) to provide users with test data for surfQuake.

*Figure 1*

## 3. Scope of the project

### 3.1 The Core Library

The core library of surfQuake is intended to automatically apply common algorithms in seismology to massive datasets. The core is written purely in Python3 and is designed to allow use with external config files or python *Dataclasses* directly. This approach makes the software flexible for users, allowing integration with their own workflows.

The core library is structured into three main processes. The first process involves generating a 'project' that serves as a guide for subsequent processes. The second process focuses on seismic event location, encompassing algorithms for automatic phase picking, associating these picks with specific events, and determining event location. The final part studies seismic source parameters through the analysis of body wave spectrums and moment tensor inversions.

Figure 2 shows a schematic of the workflow that combines all parts of the core library. The Toolboxes (Fig. 2 left panel) are organized with a master class containing methods to deliver the toolbox's functionality, and auxiliary methods to support the class structure.

*Figure 2*

### 3.2 Command Line Interface and Graphical User Interface

The Command Line Interface (CLI) serves as the intermediate programming layer designed for users seeking to automate tasks or integrate the surfQuake toolboxes into existing scripts. While the Core Library provides functionality for integration into other applications through APIs, the CLI offers advantages including efficiency, flexibility, scriptability for automation and remote accessibility.

A brief outline of the CLI structure and its functionality is displayed in Fig 3. It interacts with Core Library classes based on command line inputs or equivalently via textual commands with parameterization sourced from configuration files. SurfQuake uses INI files as configuration, chosen for their human-readable format and ease of modification by users.

The Graphical User Interface of surfQuake offers users a visually intuitive platform for interacting with the application and executing complex workflows through graphical components. It seamlessly integrates all toolboxes, providing users with easy access to various functionalities. Additionally, it includes a dedicated framework specifically designed for database exploration, enhancing the usability and versatility of the software.

*Figure 3*

### 3.3 Results storage: Database and Seismic Catalog

surfQuake utilizes an SQLite database to store focal parameters of seismic events, including CMT inversion and source spectrum parameters results. To enhance user interaction and facilitate database management, the dedicated Graphical User Interface (GUI) is powered by SQLalchemy. This specialized GUI empowers users to effortlessly execute queries, offering the flexibility to filter the database based on research region or specific time periods. The query results are not only presented in a tabular format but are also seamlessly visualized on an interactive map. Users can manipulate and explore the chosen data directly through the table, providing an intuitive and dynamic experience.

Furthermore, surfQuake incorporates a framework for statistical analysis, allowing users to perform basic analyses such as cumulative magnitude histograms or fitting the results with the Gutenberg-Richter law. This framework provides valuable insights into the seismicity data. For a comprehensive understanding of the database structure, a schematic representation of the database tables is provided in the supplementary material.

surfQuake employs a dual strategy to store results from various toolboxes by utilizing ObsPy's seismic catalog object (Krischer et al., 2015). Enhanced capabilities include generating a seismic catalog from event summary files, seamlessly integrating information from source spectrum analyses and moment tensor inversions. Dedicated Python methods facilitate efficient querying of the seismic catalog, contributing to overall accessibility and usability. Additionally, a method has been introduced for converting the catalog into a human-readable format, enhancing the user-friendly nature of surfQuake's data management. Figure 4 shows the dedicated Database-Statistics GUI displaying the dataset used to test this work. Figure 4b shows the framework included in the toolbox to display basic statistics over the chosen subset of earthquakes such as event histograms and plotting the Guttenber-Richter law.

*Figure 4*

## 4. Toolboxes Description

### 4.1 Project

In the context of surfQuake the project is a python object that contains a dictionary attribute with paths to valid seismograms files and relevant metadata information such as the sampling rate, trace start time and trace end time. This approach enables queries to efficiently search for files based on specific combinations of networks, stations, and channels, as well as within defined time spans. The other advantage of this approach is that this method eliminates the necessity for seismogram files to adhere to a specific folder structure like the common SDS structure, heritage from the well-known seismological software package SeisComP (Helmholtz-Centre Potsdam - GFZ German Research Centre for Geosciences and Gempa GmbH, 2008) structure. This alleviates the need for organizing seismogram files into a specific folder structure and potentially reduces the memory usage.

The project object has two *magic* methods. The first permits the exploration within the project itself, while the other enables the concatenation of project objects. This is very useful when users wish to extend the analysis of one dataset to another. Furthermore, the project object offers a help method of user assistance and project-specific information.

### 4.2 Phase Picking and Associator

To detect and classify P and S-wave arrivals, surfQuake uses *PhaseNet* (Zhu and Beroza, 2018) which is based on U-Nets trained on both spectrograms and time series. PhaseNet was selected because of the favorable balance between computational efficiency and phase detection rate (García et la., 2022; Münnchmeyer et al., 2022) when compared to other well-known phase pickers based in Deep Learning such as *Generalized Phase Detection* (GPD, Ross et al., 2018) and *EQTransformer* (Mousavi et al., 2020). Additionally, PhaseNets compatibility with other common Python3 numerical computation libraries Numpy (Harris et al., 2020), Scipy (Virtanen et al., 2020) and Obspy (Krischer et al., 2015) facilitates seamless integration.

PhaseNet has been integrated into the core library to easily read the project object and generate output for the associator module. Its class is specifically designed to efficiently incorporate inputs from the GUI.

Following the earthquake-monitoring workflow, the associator algorithm separates the phase arrival times into different groups corresponding to their respective events. We have chosen REAL (Zhang et al., 2019) for its simplicity, adaptable parameter configurations (eg., azimuthal coverage and weighting or automatic outlier removal), and ability to isolate events. REAL associates arrival times with events through a grid-search approach based on pick counts and calculated travel-time residuals.

In surfQuake the main class REAL contains all parametrization and transforms the file format output of Phasenet into REAL phase picking input. The parameter settings are also accessible through a dedicated and simplified configuration file, an approach particularly useful for CLI users. The specific CLI inputs for running the surfQuake associator are shown in Figure 3 and Figure 5 displays the GUI during phase picking and event association.

*Figure 5*

### 4.3 Event Location

The event location toolbox provides the capability to locate seismic events from the output of the associator. The algorithm implemented in surfQuake is the widely used nonlinear location (*NLL*) algorithm (Lomax et al., 2000). *NLL* uses the Eikonal finite-difference algorithm implemented by Podvin & Lecomte (1991) to estimate the travel times and provide probabilistic solutions through the equivalent differential time (Zhou, 1994; Font et al., 2004). In comparison with other commonly used linear event location algorithms, the solutions obtained using *NLL* are more robust in complex velocity models (Cabieces et al., 2020a). *NLL* also allows event location using either a 1D or 3D Earth velocity model based on direct search methods such as the stochastic Metropolis-Gibbs (Sambridge & Mosegaard, 2002) or the Oct-tree sampling algorithm (Lomax et al., 2009).

The locations are automatically processed for the previously associated events. A summary catalog is then generated, and results are available to be incorporated into the project database.

### 4.4 Source Spectrum Analysis

After computing a probabilistic event location, source parameters estimation can constrain further processes associated with the fault rupture and the thermal state of the crust and upper mantle. surfQuake incorporates SourceSpec (Satriano 2023), a set of command line tools, with minor adaptations to automate event computations and to enable it to be used by users with all programming levels.

SourceSpec inverts the P- and S-wave displacement amplitude spectra recorded across the seismic network to compute source parameters. The source parameters estimated include the seismic moment, corner frequency, radiated energy, source size, stress drop and local magnitude. Attenuation parameters t-star and quality factor Q are also calculated.

One of the main advantages of SourceSpec is the implementation of a robust waveform selection method that is based on the pre-signal event spectrum. In addition, SourceSpec offers the flexibility to choose different inversion algorithms including truncated Newton, Levenberg-Marquardt and basin-hopping algorithms.

Figure 6 shows an example of the typical source spectral analysis for the biggest magnitude earthquakes in the test dataset.

### 4.5 Moment Tensor Inversion

surfQuke handles methods for automated centroid moment tensor (CMT) inversion within a Bayesian framework. The functionality of the CMT toolbox is based on the Bayes-ISOLA code (Vackář et al. 2017), which performs full-waveform inversions in a space-time grid. This process is combined with analytical inversion methods to accelerate the overall procedure, centered around provided hypocenter locations. The inversion approach includes the option to calculate a data covariance matrix from pre-event noise to yield an automatic weighting function for the receiver components according to their noise levels. This also serves as an automated frequency filter to suppress noisy frequency ranges.

The results provided by the toolbox contain the best-fit solution, as well as the full posterior probability density function (PPDF), which allows the user to plot the marginal probability density functions (PDFs) for any of the CMT parameters. This method provides a best-fitting CMT solution and a full, non-Gaussian PPDF (Sambridge & Mosegaard, 2002; Sambridge, 2013).

surfQuake allows the computation of CMTs for large datasets with the corresponding event information saved in the database or the catalog. surfQuake utilities allow filtering the catalog by space-time and magnitude prior to the CMT inversion It automatically selects the corresponding waveforms from the project and computes the Green Functions to prepare the inversion. After completing the CMT inversions it saves individual results and produces an overall summary.

Figure 7 shows an example of a simple Python 3 script to run the CMT estimation for a set of events. The second part of the script includes the code for generating a summary that can be subsequently incorporated to a seismic catalog object or to the database.

## 5. A Comprehensive Test Analysis

After executing surfQuake with the test dataset, we automatically detected, associated, and located 2046 events within the designated study area. Utilizing a 90% confidence level, we calculated the mean depth and the mean major and minor axis lengths of the uncertainty ellipse for the epicenters, yielding values of 3.0 km, 3.5 km, and 1.8 km, respectively.

We successfully estimated the focal parameters (seismic moment, corner frequency, radiated energy, source size, stress drop and local magnitude) and the attenuation parameters for 945 events. Additionally, we computed the Focal Mechanism for the two events with a moment magnitude greater than 3.5 using the Bayesian Moment Tensor Inversion.

Compared to the seismic catalog of the ICGC (https://www.icgc.cat/es), our automated process enabled the detection, association, and location of twice as many events. This significant enhancement allows for the exploration and analysis of a substantially larger dataset, providing valuable insights and allowing novel findings for researchers. Moreover, we provide valuable insights into the source parameters, particularly through the estimation of the moment magnitudes along with their associated uncertainties.

The focal mechanisms retrieved by surfquake for the two main events align with those reported by the ICGC. The Moment Tensor Inversion was automatically calculated in the frequency band [0.02 - 0.30] Hz using the nearest stations to the corresponding epicenters (up to 10 stations), incorporating all components. To ensure data quality, a signal-to-noise ratio of 5 was imposed to reject low-quality seismograms. Moreover, a pre-event noise window of 20 minutes, necessary to estimate the covariance matrix, was utilized to perform a robust inversion.

Figure 8 presents the results of the seismicity location conducted in this study, revealing two primary clusters of epicenters along with their corresponding focal mechanism solutions. For additional insights into the Moment Tensor Inversion process, a detailed report is available in the Supplementary Material.

We have included a repository (see Data and Resources section) containing a subset of the dataset (see Data and Resources section), specifically focusing on the largest magnitude event, to facilitate users in familiarizing themselves with the software. This repository includes Python code examples utilizing the surfQuake core library and the command-line interface (CLI), to showcase the complete software workflow with a small dataset. Additionally, users can replicate this workflow using the graphical user interface (GUI) provided.

Our comprehensive analysis allows for the extraction of focal mechanisms and essential focal parameters such as seismic moment, corner frequency, radiated energy, source size, stress drop, and local magnitude. This extensive dataset provides researchers with a deeper understanding of seismic events, facilitating more accurate assessments of fault behavior, seismicity patterns, and tectonic processes in this region.

## 6. Future Development

The current design of surfQuake enables software developers to easily upgrade the core library. Within the context of this work, surfQuake primarily focuses on automating seismic monitoring and analyzing source parameters. While significant effort has been dedicated to providing a comprehensive toolbox, there is inevitably an ongoing necessity to enhance the core with new algorithm implementations. Looking ahead, we have outlined plans to incorporate two additional development branches in the foreseeable future.

The first planned development is the implementation of a wrapper to estimate relative locations. Integrating Relative Locations methods into the software will enhance its location capabilities, enabling a greater precision in determining the relative positions of proximate earthquakes. To this end, a particularly noteworthy addition is the integration of *NLL-SSST-COHERENCE* (Lomax & Savvaidis, 2021) to surfQuake.

The *NLL-SSST-COHERENCE* approach achieves precision comparable to differential timing while leveraging the probabilistic global-search *NLL* location method already integrated into surfQuake. A key advantage of incorporating *NLL-SSST-COHERENCE* lies in its robustness with complex earth velocity models and its capacity to retrieve precise locations even with sparse networks.

The second branch involves the development of a module to automatically estimate the slowness vector (Cabieces et al., 2020b) and locate events using seismic array techniques. By introducing this toolbox, we aim to significantly enhance the seismic monitoring capabilities of conventional networks, allowing users to harness advanced seismic array techniques for enhanced accuracy in event localization and signal quality enhancement. To accomplish this, a dedicated core toolbox and graphical user interface (GUI) are currently in development.

The toolbox will incorporate several advanced techniques for estimating the slowness vector and improving event localization. These include the conventional beamforming and frequency-wavenumber technique (Capon et al., 1967; Capon et al., 1968), which provide effective methods for estimating the slowness vector. Additionally, advanced direction-finding algorithms based on eigenvalue decomposition will be integrated, such as Multiple Signal Classification (MUSIC, Schmidt 1986) and Estimation Signal Parameters via Rotational Invariance Technique (ESPRIT, Roy 1989). These algorithms are widely used in large arrays and Distributed Acoustic Sensing (DAS) cables (van den Ende & Ampuero, 2021).

## 7. Conclusions

surfQuake offers a comprehensive suite of toolboxes designed to enhance the efficiency of seismic source parameter estimation workflows. Currently, surfQuake automates crucial tasks in seismic monitoring including arrival time determination, event localization, magnitude assessment, attenuation analysis, and moment tensor inversion.

One of the standout features of surfQuake is its adaptability, offering users three distinct programming levels for flexibility and customization. The Core Library provides functionality for integration into other applications via APIs, enabling proficient coders to build custom solutions. The Command Line interface offers stand-alone tools with direct user interaction via textual commands, and additional capabilities for scripting and automation. Additionally, the GUI software supplies a visually intuitive interface for user-friendly interaction and managing complex workflows. Each programming level caters to different use cases and user preferences, ensuring a tailored experience for diverse seismic analysis needs.

Finally, surfQuake has been successfully tested with a one year of data from the central Pyrenees region, proving its capability to automatically determine source parameters from over 2000 events. surfQuake stands as a reliable and efficient solution for seismic analysis tasks.

**Data and Resources**

surfQuake is freely available at https://projectisp.github.io/surfquaketutorial.github.io/. A minimal real-data example and tutorial is also available at https://github.com/rcabdia/test_surfquake.

Waveform data used to test surfQuake were retrieved from the Institut Cartogràfic i Geològic de Catalunya (CA/ICGC: https://doi.org/10.7914/SN/CA) and the French Seismological and Geodetic (FR/RESIF: https://doi.org/10.17616/R37Q06) network surfQuake makes use of the ObsPy library (Krischer et al., 2015) and the graphical user interfaces are built upon the PyQt5 library (https://www.riverbankcomputing.com, last accessed November 2021). The graphics are plotted using Matplotlib (Hunter, 2007), and the maps are generated using Cartopy (https://scitools. org.uk/cartopy/, last accessed Abr 2024).

Other relevant data and resources were obtained from the following websites: MkDocs (https://www.mkdocs.org/, last accessed Apr 2024), SQLAlchemy (https://www.sqlalchemy. org/, last accessed Apr 2024)

**Declaration of Competing Interests**

The authors acknowledge that there are no conflicts of interest recorded.

**Data and Resources**

We thank the UPFLOW project, funded by the European Research Council under the European Union's Horizon 2020 research and innovation program (grant agreement No 101001601).

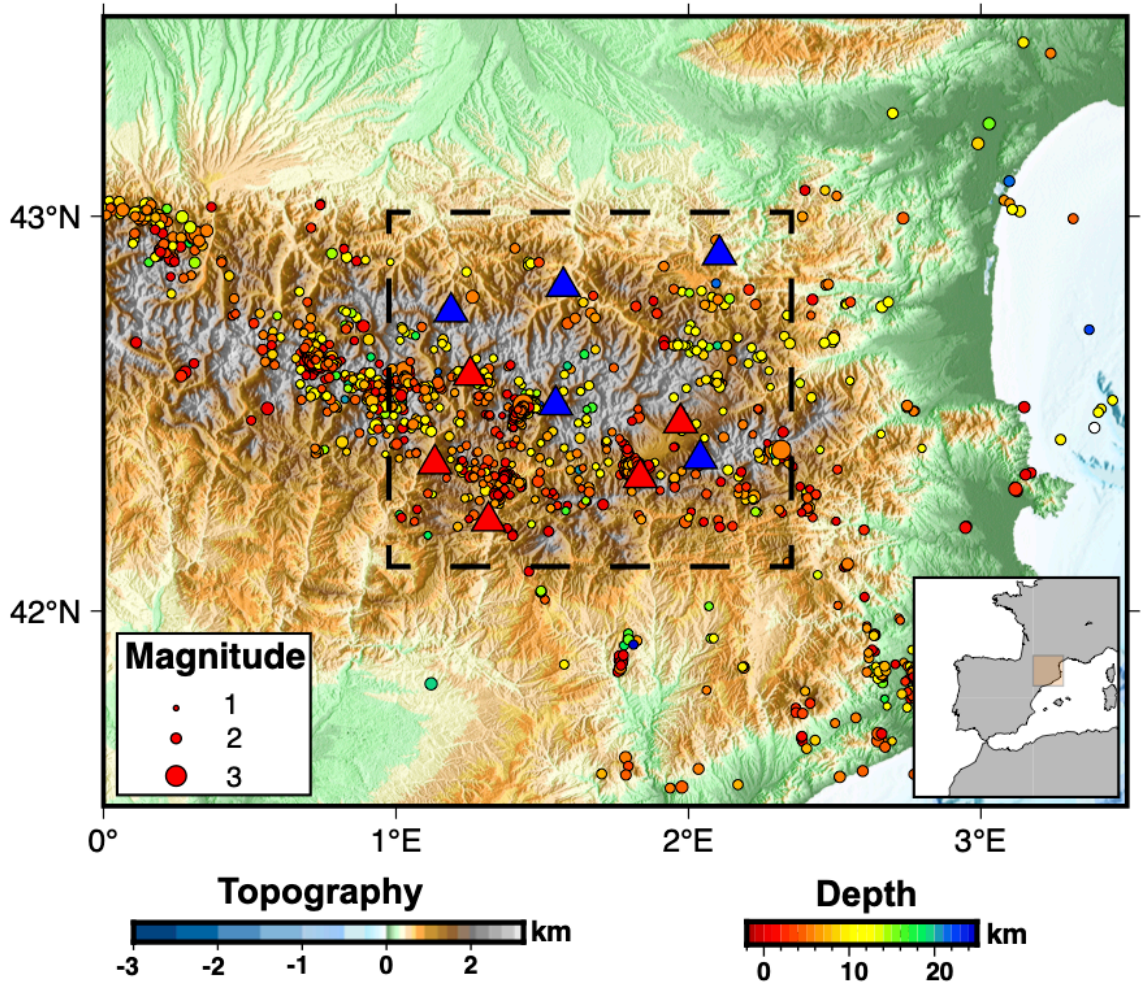

*Figure 1. Region of study with the station distribution used in the software test. Red triangles denote seismic stations from the CA/ICGC network. Blue triangles represent seismic stations from the FR/RESIF network. Circles represent hypocenters located in the ICGJ seismic catalog between Oct 2021 and Oct 2022.*

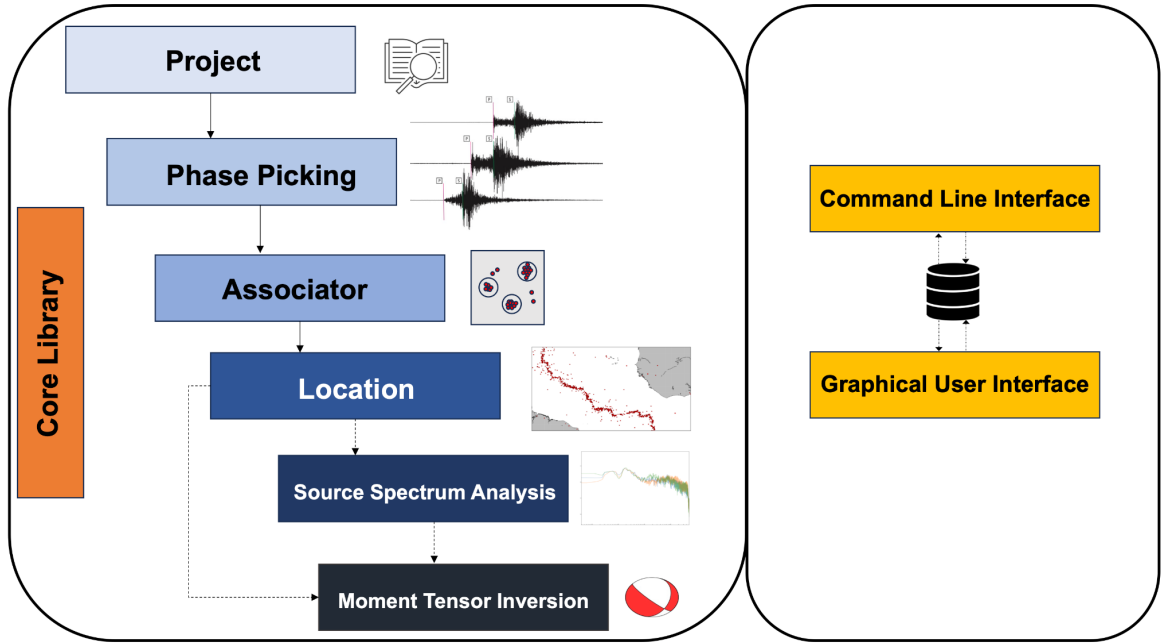

*Figure 2. Workflow diagram. Left panel illustrates the core Library, describing the core toolboxes contained inside surfQuake. Right panel illustrates the Command Line Interface and Graphical User Interface connected with the result storage system. The CLI utilizes an ObsPy catalog object, while the GUI operates on a SQLite database managed by a SQLAlchemy object-relational mapper and a dedicated graphical framework.*

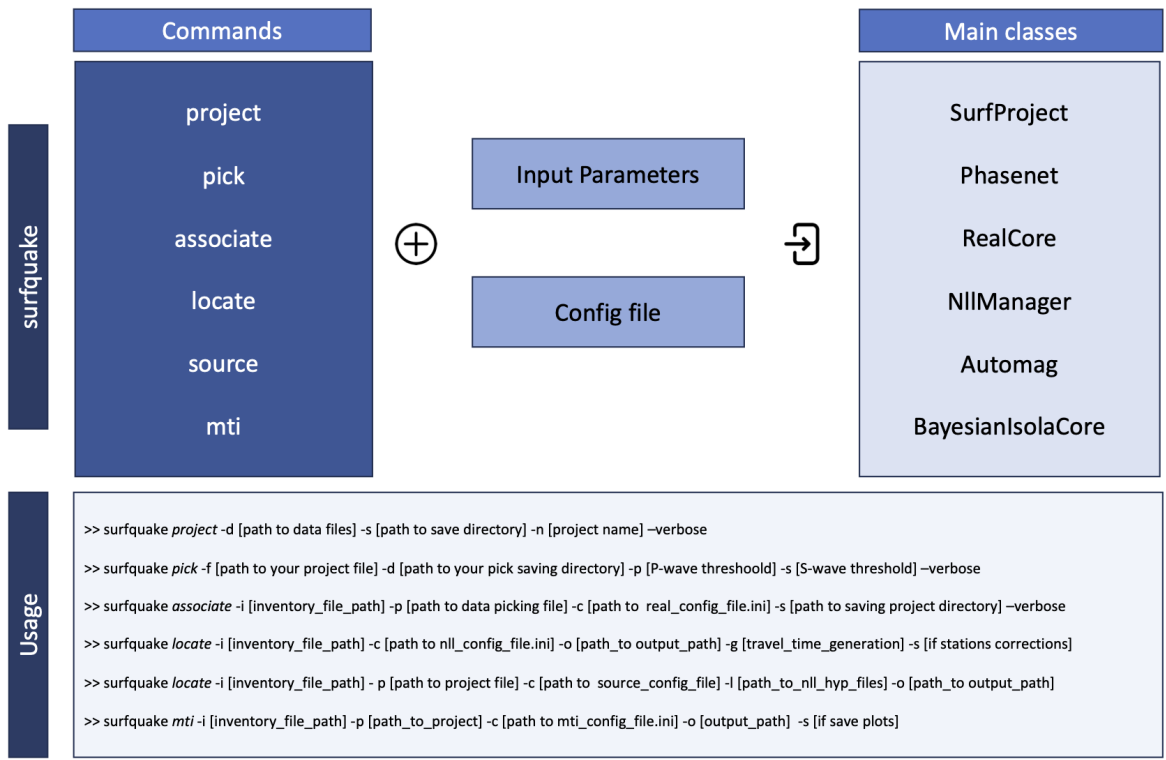

*Figure 3. Command Line Interface scheme. The left panel presents the available commands to call the Core Library toolboxes. Each command requires a specific input parameterization, represented in the middle panel. The right side displays the Main Core Library Python classes used by the command interface. The bottom panel shows basic command examples demonstrating the usage of the command line interface.*

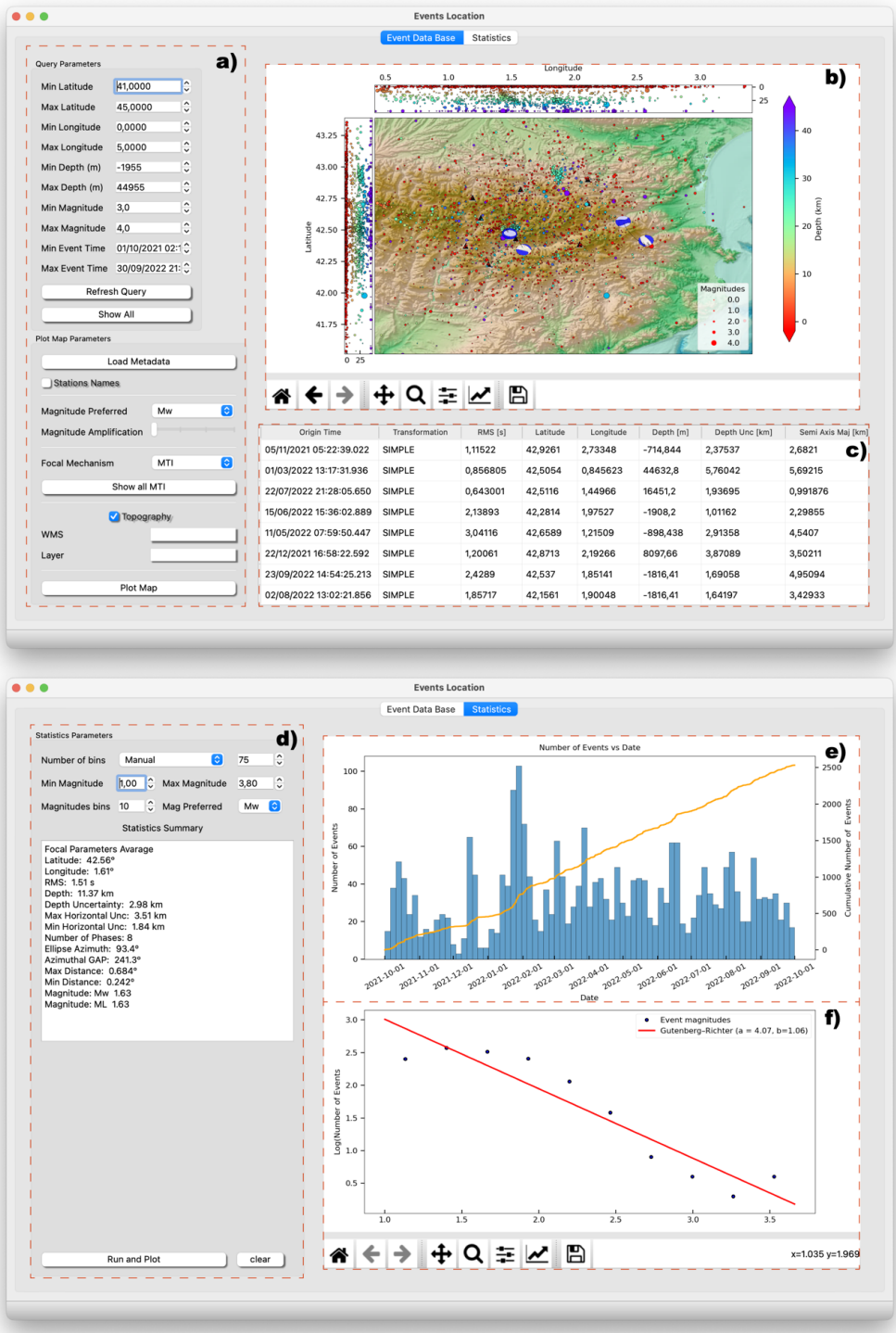

*Figure 4. Database Framework. a) Database query and mapping parametrization. b) Database map with hypocenter, stations, and Web Map Services (WMS) as available layers. Focal mechanisms can also be plotted in this widget. d) Statistics parametrization and text box showing statistics results from the database query. e) and f) display a histogram with a cumulative linear plot and a Gutenberg-Richter fit plot, respectively, of events contained in the Database query.*

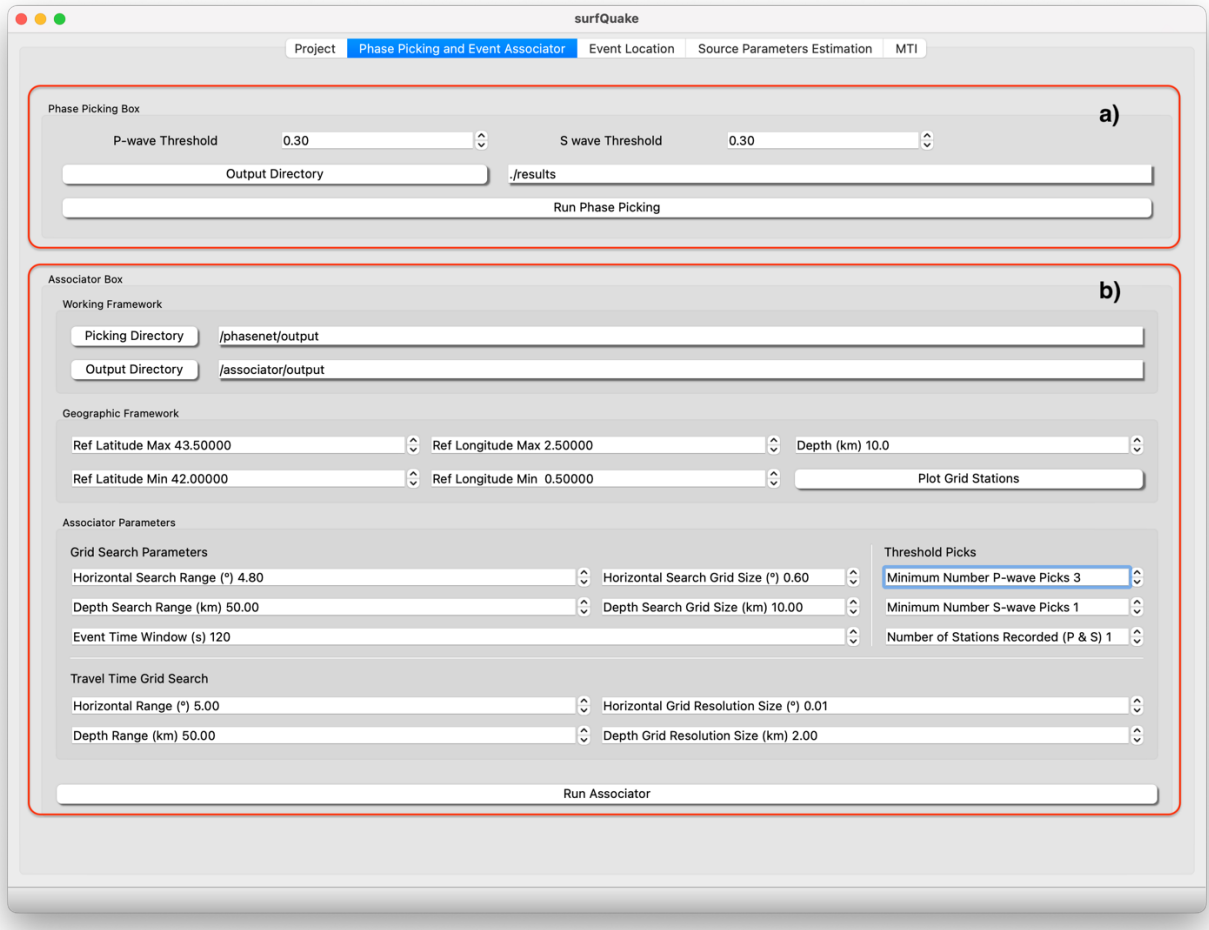

*Figure 5. GUI displaying the Phase picking and associator toolboxes. a) Dedicated parametrization widget to run the picking neural network (Phansenet, Zhu & Beroza 2019) in the previously created/loaded project. b) Dedicated parametrization widget to run the associator algorithm (REAL, Zhang et al., 2021).*

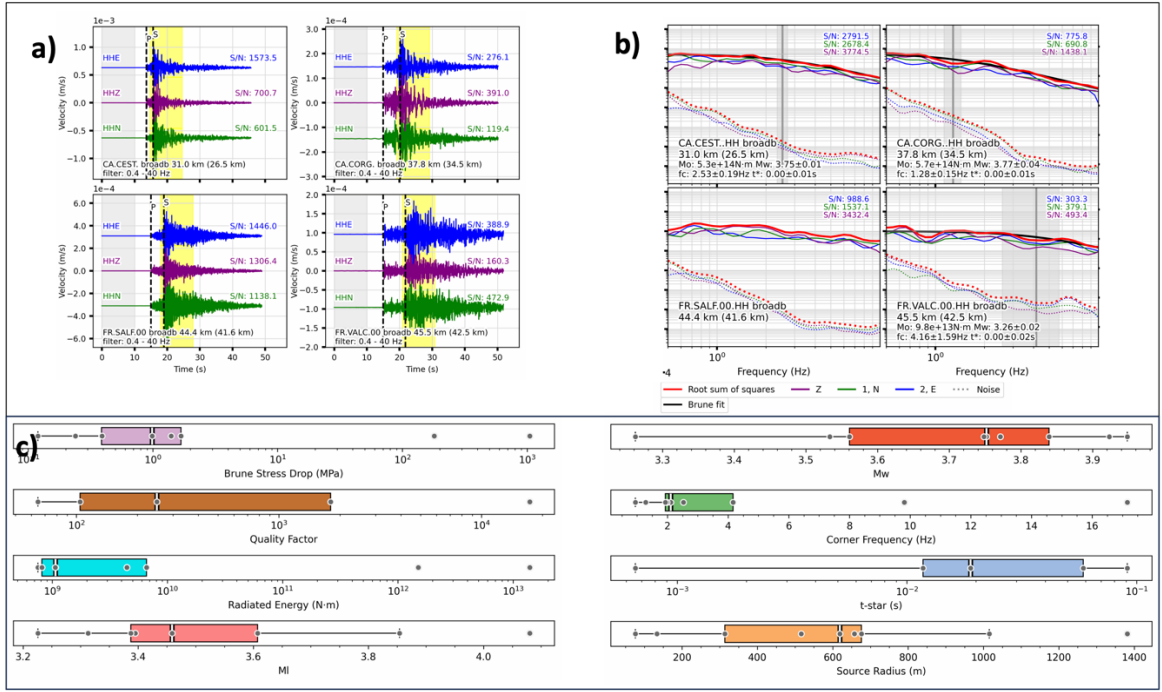

*Figure 6. Example of Source Spectrum Analysis of the selected test earthquake, 01-02-2022 / Mw 3.8. a) waveforms and S-wave time window selection to compute source spectrum inversion. b) Spectrum inversion and c) Error bars are utilized to display the results and statistics of the source spectrum parameters Mw, Mo, Radiated Energy, Corner Frequency, Source Radius, Quality Factor, Brune Stress Drop, t\* and ML.*

```python
from surfquakecore.project.surf_project import SurfProject
from surfquakecore.moment_tensor.sq_isola_tools import BayesianIsolaCore
from surfquakecore.moment_tensor.mti_parse import WriteMTI

sp = SurfProject.load_project(path_to_project_file = path_to_project_file)

bic = BayesianIsolaCore(project = sp, inventory_file = inventory_file_path,
output_directory = output_dir_path, save_plots = True)

bic.run_inversion(mti_config = config_files_path)

wm = WriteMTI(parsed_args.output_dir_path)

file_summary = os.path.join(output_dir_path, "summary_mti.txt")

wm.mti_summary(output = file_summary)
```

*Figure 7. Simple example of a Python script using the Moment Tensor Inversion Core Library toolbox. The top part of the script is calling the core library previously installed in the computer system. The bottom part of the script is loading a surfQuake project, instantiating a Bayesian Isola Core object, and then running the Moment Tensor Inversion Finally the script writes a summary with the most remarkable aspects of the inversion output information.*

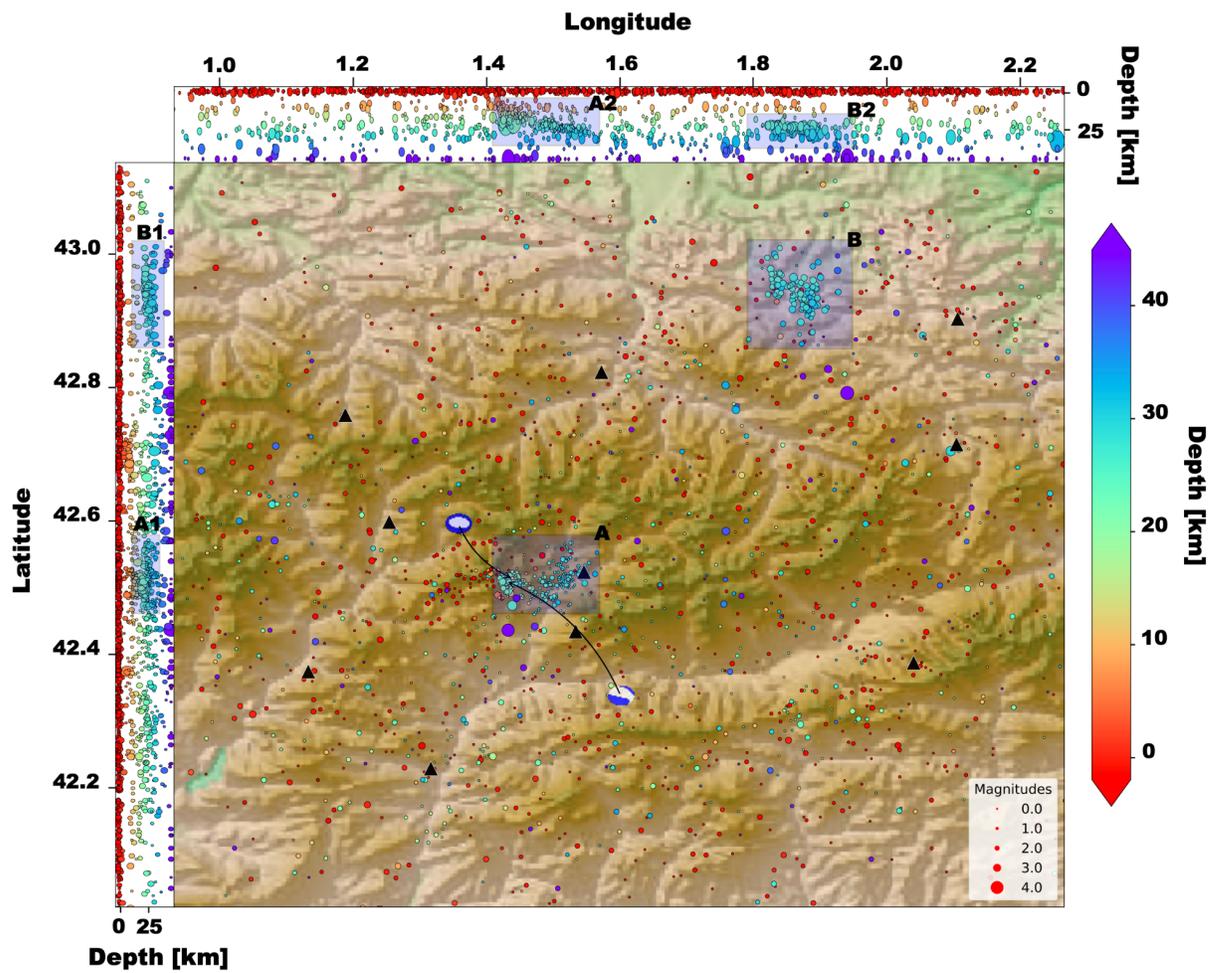

*Figure 8. Seismicity Test Results presents the Seismicity Test Results, offering a detailed view of the relocated hypocenters from the test conducted in the central Pyrenees region. Square A highlights a cluster on the map section while A1 and A2 highlight the same cluster on the cross sections. B, B1 and B2 highlight a northern cluster on the map and cross sections, respectively. Black Triangles represent seismic stations used in the event relocation test. The focal mechanisms estimated for events with magnitude bigger than 3.5 are detailed in the Supplementary Material.*

# References

Cabieces, R., Buforn, E., Cesca, S., & Pazos, A. (2020). Focal Parameters of Earthquakes Offshore Cape St. Vincent Using an Amphibious Network. *Pure and Applied Geophysics*, *177*(4), 1761–1780. https://doi.org/10.1007/s00024-020-02475-3

Cabieces, R., Krüger, F., Garcia-Yeguas, A., Villaseñor, A., Buforn, E., Pazos, A., Olivar-Castaño, A., & Barco, J. (2020). Slowness vector estimation over large-aperture sparse arrays with the Continuous Wavelet Transform (CWT): Application to Ocean Bottom Seismometers. *Geophysical Journal International*, *223*(3), 1919–1934.

Cabieces, R., Olivar-Castaño, A., Junqueira, T. C., Relinque, J., Fernandez-Prieto, L., Vackár, J., Rösler, B., Barco, J., Pazos, A., & García-Martínez, L. (2022). Integrated Seismic Program (ISP): A New Python GUI-Based Software for Earthquake Seismology and Seismic Signal Processing. *Seismological Research Letters*, *93*(3), 1895–1908.

Capon, J. (1969). High-resolution frequency-wavenumber spectrum analysis. *Proceedings of the IEEE*, *57*(8), 1408–1418.

Capon, J., Greenfield, R. J., & Kolker, R. J. (1967). Multidimensional maximum-likelihood processing of a large aperture seismic array. *Proceedings of the IEEE*, *55*(2), 192–211.

Font, Y., Kao, H., Lallemand, S., Liu, C.-S., & Chiao, L.-Y. (2004). Hypocentre determination offshore of eastern Taiwan using the Maximum Intersection method. *Geophysical Journal International*, *158*(2), 655–675.

García, J. E., Fernández-Prieto, L. M., Villaseñor, A., Sanz, V., Ammirati, J.-B., Díaz Suárez, E. A., & García, C. (2022). Performance of deep learning pickers in routine network processing applications. *Seismological Society of America*, *93*(5), 2529–2542.

Goldstein, P., Dodge, D., Firpo, M., Minner, L., Lee, W., Kanamori, H., Jennings, P., & Kisslinger, C. (2003). SAC2000: Signal processing and analysis tools for seismologists and engineers. *The IASPEI International Handbook of Earthquake and Engineering Seismology*, *81*, 1613–1620.

Harris, C. R., Millman, K. J., Walt, S. J. van der, Gommers, R., Virtanen, P., Cournapeau, D., Wieser, E., Taylor, J., Berg, S., Smith, N. J., Kern, R., Picus, M., Hoyer, S., Kerkwijk, M. H. van, Brett, M., Haldane, A., Río, J. F. del, Wiebe, M., Peterson, P., … Oliphant, T. E. (2020). Array programming with NumPy. *Nature*, *585*(7825), 357–362. https://doi.org/10.1038/s41586-020-2649-2

Havskov, J., Voss, P. H., & Ottemöller, L. (2020). Seismological observatory software: 30 Yr of SEISAN. *Seismological Research Letters*, *91*(3), 1846–1852.

Heimann, S., Vasyura-Bathke, H., Sudhaus, H., Isken, M. P., Kriegerowski, M., Steinberg, A., & Dahm, T. (2019). A Python framework for efficient use of pre-computed Green's functions in seismological and other physical forward and inverse source problems. *Solid Earth*, *10*(6), 1921–1935.

Helmholtz-Centre Potsdam - GFZ German Research Centre for Geosciences and gempa GmbH. (2008). *The SeisComP seismological software package. GFZ Data Services.* https://doi.org/10.5880/GFZ.2.4.2020.003

Herrmann, R. B. (2013). Computer programs in seismology: An evolving tool for instruction and research. *Seismological Research Letters*, *84*(6), 1081–1088.

Krischer, L., Megies, T., Barsch, R., Beyreuther, M., Lecocq, T., Caudron, C., & Wassermann, J. (2015). ObsPy: A bridge for seismology into the scientific Python ecosystem. *Computational Science & Discovery*, *8*(1), 014003.

Lomax, A., Michelini, A., Curtis, A., Meyers, R., & others. (2009). Earthquake location, direct, global-search methods. *Encyclopedia of Complexity and Systems Science*, *5*, 2449–2473.

Lomax, A., Virieux, J., Volant, P., & Berge-Thierry, C. (2000). Probabilistic earthquake location in 3D and layered models. In *Advances in seismic event location* (pp. 101–134). Springer.
24

Lomax, A., & Savvaidis, A. (2022). High-precision earthquake location using source-specific station terms and inter-event waveform similarity. *Journal of Geophysical Research: Solid Earth*, *127*(1), e2021JB023190.

Mousavi, S. M., Ellsworth, W. L., Zhu, W., Chuang, L. Y., & Beroza, G. C. (2020). Earthquake transformer—An attentive deep-learning model for simultaneous earthquake detection and phase picking. *Nature Communications*, *11*(1), 3952.

Münchmeyer, J. (2024). PyOcto: A high-throughput seismic phase associator. *Seismica*, *3*(1). https://doi.org/10.26443/seismica.v3i1.1130

Podvin, P., & Lecomte, I. (1991). Finite difference computation of traveltimes in very contrasted velocity models: A massively parallel approach and its associated tools. *Geophysical Journal International*, *105*(1), 271–284.

Ross, Z. E., Meier, M.-A., Hauksson, E., & Heaton, T. H. (2018). Generalized seismic phase detection with deep learning. *Bulletin of the Seismological Society of America*, *108*(5A), 2894–2901.

Roy, R., & Kailath, T. (1989). ESPRIT-estimation of signal parameters via rotational invariance techniques. *IEEE Transactions on Acoustics, Speech, and Signal Processing*, *37*(7), 984–995.
25

Sambridge, M. (2014). A parallel tempering algorithm for probabilistic sampling and multimodal optimization. *Geophysical Journal International*, *196*(1), 357–374.

Sambridge, M., & Mosegaard, K. (2002). Monte Carlo methods in geophysical inverse problems. *Reviews of Geophysics*, *40*(3), 3–1.

Satriano, C. (2023). *SourceSpec – Earthquake source parameters from P- or S-wave displacement spectra (X.Y)*. https://doi.org/10.5281/ZENODO.3688587

Schmidt, R. (1986). Multiple emitter location and signal parameter estimation. *IEEE Transactions on Antennas and Propagation*, *34*(3), 276–280.

Vackář, J., Burjánek, J., Gallovič, F., Zahradník, J., & Clinton, J. (2017). Bayesian ISOLA: New tool for automated centroid moment tensor inversion. *Geophysical Journal International*, *210*(2), 693–705. https://doi.org/10.1093/gji/ggx158

Si, X., Wu, X., Li, Z., Wang, S., & Zhu, J. (2024). An all-in-one seismic phase picking, location, and association network for multi-task multi-station earthquake monitoring. *Communications Earth & Environment*, *5*(1), 22.

van den Ende, M. P., & Ampuero, J.-P. (2021). Evaluating seismic beamforming capabilities of distributed acoustic sensing arrays. *Solid Earth*, *12*(4), 915–934.

Virtanen, P., Gommers, R., Oliphant, T. E., Haberland, M., Reddy, T., Cournapeau, D., Burovski, E., Peterson, P., Weckesser, W., Bright, J., van der Walt,

S. J., Brett, M., Wilson, J., Millman, K. J., Mayorov, N., Nelson, A. R. J., Jones, E., Kern, R., Larson, E., … SciPy 1.0 Contributors. (2020). SciPy 1.0: Fundamental Algorithms for Scientific Computing in Python. *Nature Methods*, *17*, 261–272. https://doi.org/10.1038/s41592-019-0686-2

Wu, X., Huang, S., Xiao, Z., & Wang, Y. (2022). Building Precise Local Submarine Earthquake Catalogs via a Deep-Learning-Empowered Workflow and its Application to the Challenger Deep. *Frontiers in Earth Science*, *10*, 817551.

Zhang, M., Ellsworth, W. L., & Beroza, G. C. (2019). Rapid Earthquake Association and Location. *Seismological Research Letters*, *90*(6), 2276–2284. https://doi.org/10.1785/0220190052

Zhang, M., Liu, M., Feng, T., Wang, R., & Zhu, W. (2022). LOC-FLOW: An end-to-end machine learning-based high-precision earthquake location workflow. *Seismological Society of America*, *93*(5), 2426–2438.

Zhang, X., Zhang, M., & Tian, X. (2021). Real-time earthquake early warning with deep learning: Application to the 2016 M 6.0 Central Apennines, Italy earthquake. *Geophysical Research Letters*, *48*(5), 2020GL089394.

Zhou, H. (1994). Rapid three-dimensional hypocentral determination using a master station method. *Journal of Geophysical Research: Solid Earth*, *99*(B8), 15439–15455.

Zhu, W., & Beroza, G. C. (2019). PhaseNet: A deep-neural-network-based seismic arrival-time picking method. Geophysical Journal International, 216(1), 261–273. https://doi.org/10.1093/gji/ggy423

Zhu, W., McBrearty, I. W., Mousavi, S. M., Ellsworth, W. L., & Beroza, G. C. (2022). Earthquake phase association using a Bayesian Gaussian mixture model. Journal of Geophysical Research: Solid Earth, 127(5), e2021JB023249.

*Supplementary Material*

## *surfQuake: A new Python toolbox for the workflow process of seismic sources*

Roberto Cabieces[1], Thiago C. Junqueira[2], Katrina Harris[3], Jesús Relinque[1], Claudio Satriano[4].

1. Department of Geophysics, Spanish Navy Observatory, San Fernando, Spain.

2. Institute of Geoscience, University of Potsdam, Karl-Liebknecht-Str., 24–2514476 Potsdam-Golm, Germany.

3. Department of Earth Sciences, University College London, Gower Place, WC1H6BT London, UK.

4. Institut de Physique du Globe de Paris (IPGP), 1 Rue Jussieu 75238 Paris cedex 05, France

**Table A1.** Seismic stations information

| Net.Station | Sensor | Digitizer |
|---|---|---|
| **CA.ARBS** | CMG3TB | Centaur (Nanometrics) |
| **CA.CEST** | CMG3T | Centaur (Nanometrics) |
| **CA.CORG** | STS2.5 | Centaur (Nanometrics) |
| **CA.CSOR** | STS2 | Centaur (Nanometrics) |
| **FR.GENF** | TRILLIUM-120PH | Centaur (Nanometrics) |
| **FR.CARF** | TRILLIUM-120QA | D6BB (Staneo) |
| **FR.FNEB** | TRILLIUM-120PA | D6BB (Staneo) |
| **FR.PAND** | TRILLIUM-Compact | D6BB (Staneo) |
| **FR.SALF** | TRILLIUM-120QA | D6BB (Staneo) |
| **FR.VALC** | TRILLIUM-120PH | Centaur (Nanometrics) |

## A1. Seismic Moment Tensor Solution

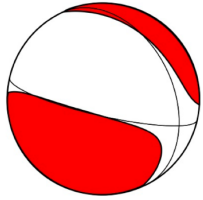

Origin time: 2022-02-01 02:02:58
Lat  42.510   Lon   1.428   Depth 20.9 km

Estimated rupture length:  0.665 km
VR:  25 %

MT [ Mrr   Mtt   Mpp   Mrt   Mrp   Mtp ]:
  [-1.83   1.61   0.22  -3.15   0.61  -1.81 ] * 1e+14

Scalar Moment: M0 =  4.33e+14 Nm (Mw = 3.7)

DC component:  73 %  CLVD component:  27 %
ISOtropic component:  0 %

Fault plane 1: strike =  334, dip =  25 , slip-rake = -42
Fault plane 2: strike =  103, dip =  74 , slip-rake = -109

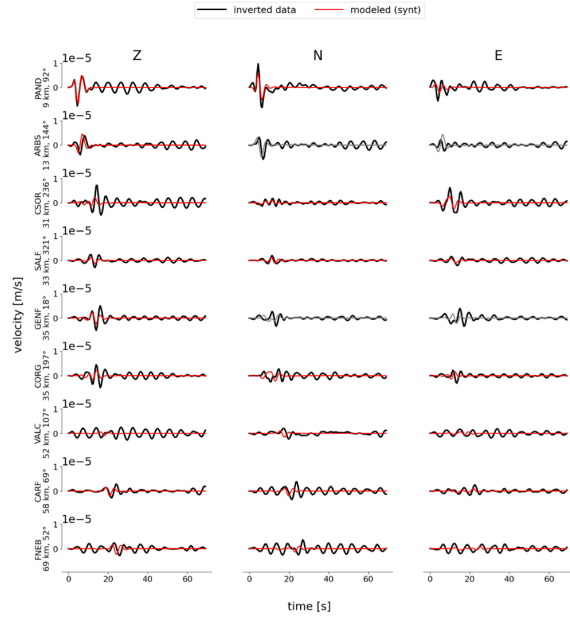

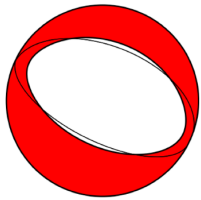

Origin time: 2021-10-11 08:23:14
Lat  42.513   Lon   1.441   Depth 19.6 km

Estimated rupture length:  0.471 km
VR:  17 %

MT [ Mrr   Mtt   Mpp   Mrt   Mrp   Mtp ]:
  [ 1.92   4.87   3.73  -0.44   0.17  -0.18 ] * 1e+14

Scalar Moment: M0 =  5.17e+14 Nm (Mw = 3.8)

DC component:  24 %  CLVD component: -8 %
ISOtropic component:  68 %

Fault plane 1: strike =  276, dip =  36 , slip-rake = -95
Fault plane 2: strike =  102, dip =  54 , slip-rake = -86

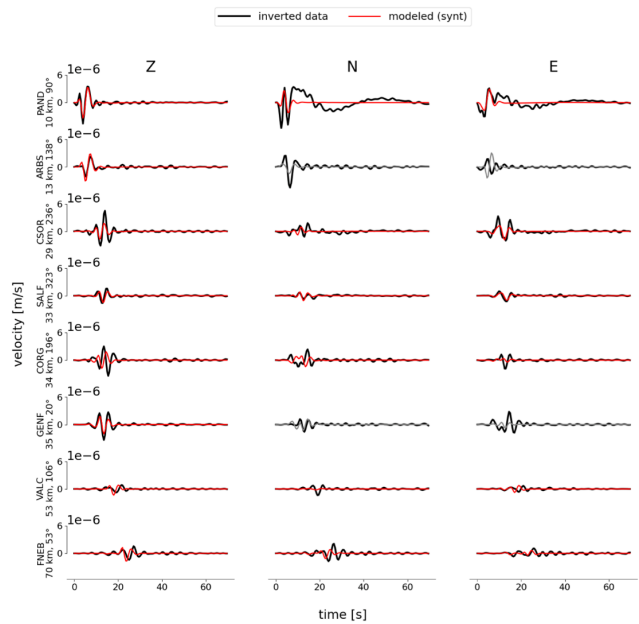

## A2. Database Structure

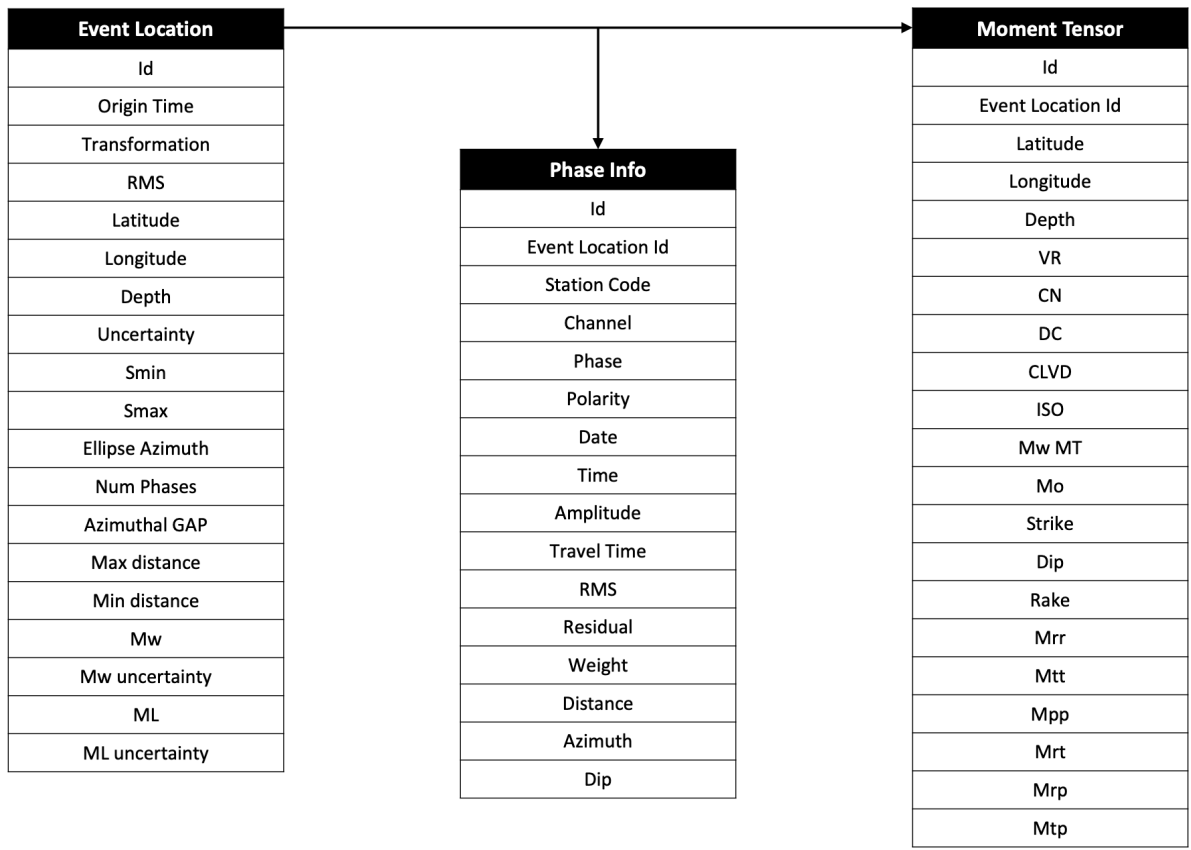